\documentclass[twocolumn]{aastex6}
\pdfoutput=1 
\usepackage{amsmath,amsthm,amssymb,amsfonts}
\usepackage{float,xspace,makecell}
\usepackage[figuresright]{rotating}
\usepackage{placeins}

\newcommand\mosfit{{\tt MOSFiT}\xspace}

\shorttitle{An Energy Inventory of Tidal Disruption Events}
\shortauthors{Mockler \& Ramirez-Ruiz}

\begin{document}

\title{An Energy Inventory of Tidal Disruption Events}

\author{Brenna Mockler\altaffilmark{1,2} and Enrico Ramirez-Ruiz\altaffilmark{1,2}}
\affil{$^1$Department of Astronomy and Astrophysics, University of California, Santa Cruz, CA 95064, USA\\
$^2$DARK, Niels Bohr Institute, University of Copenhagen, Blegdamsvej 17, 2100 Copenhagen, Denmark}

\email{bmockler@ucsc.edu}

\begin{abstract}
Tidal disruption events (TDEs) offer a unique opportunity to study a single super-massive black hole (SMBH) under feeding conditions that change over timescales of days or months. However, the primary mechanism for generating luminosity during the flares remains debated. Despite the increasing number of observed TDEs, it is unclear whether most of the energy in the initial flare comes from accretion near the gravitational radius or from circularizing debris at larger distances from the SMBH. The energy dissipation efficiency increases with decreasing radii, therefore by measuring the total energy emitted and estimating the efficiency we can derive clues about the nature of the emission mechanism. Here we calculate the integrated energy, emission timescales, and average efficiencies for the TDEs using the Modular Open Source Fitter for Transients (\mosfit). Our calculations of the total energy generally yield higher values than previous estimates. This is predominantly because, if the luminosity follows the mass fallback rate, TDEs release a significant fraction of their energy long after their light curve peaks. We use \mosfit to calculate the conversion efficiency from mass to radiated energy, and find that for many of the events it is similar to efficiencies inferred  for active galactic nuclei. There are, however, large systematic uncertainties in the measured efficiency due to model degeneracies between the efficiency and the mass of the disrupted star, and these must be reduced before we can definitively resolve the emission mechanism of individual TDEs. 
\end{abstract}


\section{Introduction}\label{sec:intro}
The luminosity of supermassive black holes (SMBHs) residing in the nucleus of most, if not all, galaxies is directly related to the rate at which they are supplied with matter.  While active galactic nuclei are supplied by steady flows of fuel for thousands of years, tidal disruption events offer the  distinctive possibility of investigating a single SMBH under feeding conditions that vary over timescales of weeks or months \citep[e.g.,][]{Dai:2018}.

Nonetheless, the recent breakthroughs in observations of tidal disruption events have highlighted our incomplete theoretical understanding of these transients \citep[e.g.,][]{Hung:2017a}. The candidate flares rise and fall in brightness over a period of weeks to months, with power-law decline rates that agree (at least for the first several months) with numerical predictions of the rates at which stellar debris falls onto the SMBH \citep[e.g.,][]{Guillochon:2013a}. However, we still do not have a good understanding of how infalling material circularizes and accretes onto the SMBHs, or how or where the emission we observe is generated. Because the source of the emission appears to be obscured by an optically thick reprocessing layer throughout most of the observed light curve, our understanding of the inner processes generating the radiation remains incomplete \citep[e.g.,][]{Roth:2016a}. 

The compilation of an energy inventory  offers an overview of the integrated effects of the energy transfers involved in tidal disruption events. By estimating the efficiency of conversion of mass into electromagnetic radiation, it is possible to constrain whether the emission originates from accretion onto the black hole, or whether it originates at larger radii during the circularization of bound stellar debris. 

Observational studies have shown that the energy estimated from the observations of TDEs does not add up to the total energy expected from the accretion of half of a solar mass of material, which corresponds to $\approx 10^{53} (\epsilon/0.1) (M_{\ast} /0.5 M_{\odot})$ ergs assuming a commonly used efficiency value of $0.1$ (all but one of the events discussed in this work have energy estimates below $10^{53}$ ergs). To solve this {\it missing energy problem}, it has been suggested that if the luminosity is generated instead from circularization processes such as stream collisions, which occur at larger radii from the black hole, the mass-to-energy efficiency would decrease, lowering the total expected energy and reducing the apparent tension between theory and observations \citep{Piran:2015b}. Other papers addressing this discrepancy have suggested that a large fraction of the rest mass energy is carried by outflows \citep{Metzger:2016b}, or that additional energy is emitted either at higher frequencies that are not covered by existing observations \citep{Lu:2018}, or at later times by an accretion disk with a very long viscous timescale \citep{van-Velzen:2019a}. To aid this discussion we provide a robust calculation both of the total energy radiated during these flares and the associated conversion timescales.  

Most observational studies of optical and UV TDEs include estimates of the bolometric luminosity at peak or at discovery. Some also include estimates of the radiated energy by integrating the  bolometric light curve. For the transients discussed in this paper, the literature estimates of bolometric luminosity curves are calculated either by fitting a simple model to the observed light curve \citep{Gezari:2008a, Gezari:2012a, Hung:2017a}, applying a bolometric correction to the optical/UV light curves  \citep{Chornock:2014a}, or fitting thermal blackbody spectra to the photometric  light curves \citep{Holoien:2014a, Holoien:2016b, Holoien:2016a,  Hung:2017a, Blagorodnova:2017a}. 

While these are important steps towards estimating the total energy emitted by these flares, the data often does not have wide band coverage (many events have limited UV coverage), and generally does not extend for more than a few peak times, leading to large uncertainties in the estimates of the total energy. Here we use a different approach. We require the bolometric luminosity to follow the expected fallback rate of the stellar debris, allowing us to fit the light curve of each event using the Markov-Chain Monte Carlo (mcmc) fitting code \mosfit \citep{Guillochon:2018, Mockler:2019a}. This allows for a robust estimate of the total energy radiated to be made not only while observations were taken, but also at early and late times before the transient was discovered and after the observing campaign ended. 
The inventory, which is presented in Table~\ref{table:Lptp}, is arranged by individual events and components within energy and timescale categories, and includes energy estimates from the literature. The explanations for each entry are presented in Section~\ref{sec:calorimetry}. A guide on how  the conversion efficiency of mass into electromagnetic radiation is estimated is detailed in Section~\ref{sec:efficiency}. Finally, Section~\ref{sec:disc} discusses the implications of the compilation and how it offers a way to assess how well we understand the physical processes at play.

\section{Calorimetry}\label{sec:calorimetry}
To constrain the energy released in optical and UV tidal disruption events, we assume that the bolometric luminosity follows the mass fallback rate and fit an evolving blackbody to the optical and UV light curves.  The assumption that the luminosity follows the fallback rate works well for the light curves of many optical and UV TDEs \citep{Mockler:2019a}, and is the expected result if circularization is prompt and an accretion disk is formed on timescales less than the fallback timescale or if the emission is produced by circularization processes such as stream collisions \citep[e.g.,][]{Bonnerot:2016a}. 

By fitting a dynamic blackbody photosphere to the optical and UV bands we are approximating the emission as efficiently thermalized. This is the simplest approximation we can make for the reprocessing of the fallback luminosity, and because optically large emitting bodies can not generally outperform a blackbody of the same size when radiating into free space, this provides us with a robust estimate on the total radiated energy. The estimates of the total bolometric energy calculated using our model can be found in Table~\ref{table:Lptp}. 
For this work we have re-run the fits presented in \citet{Mockler:2019a} with an expanded efficiency range, allowing the minimum efficiency to go down to $5\times 10^{-4}$, to ensure we are not biasing our results to higher efficiency estimates. This additional freedom does not significantly change our estimates of the integrated energy, however it does change the median values for the efficiency and stellar mass parameters. We discuss this further in Section~\ref{sec:efficiency}.

We compare the total bolometric luminosity calculated with our model ($\rm E_{bol}$) to $L_{\rm peak} \times t_{\rm peak}$. Transient events are commonly characterized in the $L_{\rm peak} - t_{\rm peak}$ plane,
and $L_{\rm peak} \times t_{\rm peak}$ seems like a natural way to approximate the energy released by a transient. We find that in most cases $L_{\rm peak} \times t_{\rm peak}$ is less than 1/3 of the total emitted energy for TDEs. Therefore, we caution against using $L_{\rm peak} \times t_{\rm peak}$ when estimating the total energy released for TDEs. We note that throughout this paper we define $t_{\rm peak} =  t_{\rm peak, \; ff}$ as the $\Delta t$ between when mass first begins to return to the black hole and when the light curve peaks -- the `time of peak from first fallback'. This is less than $t_{\rm peak}$ as calculated from disruption, but can be more easily tied to observations, as mass needs to return to the vicinity of the black hole before stream collisions or accretion produce luminosity. On the other hand,  $t_{\rm peak, \; ff}$ will be larger than the peak timescale calculated using the first observation of the TDE, as our fits extrapolate back before the first observation to the approximate first fallback time using simulated mass fallback rates.

\begin{table*}
\begin{center} 
    \renewcommand{\thefootnote}{\arabic{footnote}}
    \footnotesize
    \setlength\tabcolsep{1.4pt}
    \renewcommand{\arraystretch}{1.4}
\begin{tabular}{cccccccccccccccc}
\hline
TDE & $\rm E_{\rm bol, \; \rm lit}$ & $\rm E_{\rm bol}$& $\rm E_{\rm bol}$ & $\frac{\rm E_{\rm bol} \; (obs.)}{\rm E_{\rm bol}\; (full)}$ & $\frac{\rm L_{\rm p} \times \rm t_{\rm p,ff}}{\rm E_{\rm bol}\; \rm (full)}$ & ${\rm t_{ p,ff}}$ & $\frac{\rm t_{\rm 50}}{\rm t_{\rm p,ff}}$ & $\frac{\rm t_{\rm 90}}{\rm t_{\rm p, ff}}$ & avg. $\epsilon$ & $\epsilon$ at $\rm L_{\rm p}$ & $\frac{\rm L_{\rm p}}{\rm L_{\rm Edd}}$ & $\rm \Delta M_{min}$ \\
&& (obs.\,times) & (full\,curve) &&&&&&&&&\\
& ($\rm log_{10}ergs$) & ($\rm log_{10}ergs$) & ($\rm log_{10}ergs$) & & &(days) & & & ($\rm log_{10}$) &($\rm log_{10}$) & & ($\rm log_{10}M_{\odot}$)\\
 & & (0.09\,dex) & (0.11\,dex) & & & (15\,days) & (0.24\,dex) & (0.32\,dex) & (0.68\,dex) & & (0.17\,dex) &  \\
\hline
\bf{10jh} & $\geq 51.3$ & $51.4_{-0.1}^{+0.1}$ & $51.6_{-0.1}^{+0.1}$ & $0.68_{-0.03}^{+0.02}$ & $0.32_{-0.02}^{+0.01}$ & $55_{-1}^{+1}$ & $3.3_{-0.2}^{+0.4}$ & $37.4_{-7.3}^{+27.9}$  & $-3.3_{-0.0}^{+0.1}$ & $-3.3_{-0.0}^{+0.1}$ & $0.15_{-0.02}^{+0.02}$ & $-2.7_{-0.1}^{+0.1}$ \\
D1-9 & $\geq 50.9$ & $52.3_{-0.4}^{+0.2}$ & $53.0_{-0.4}^{+0.2}$ & $0.21_{-0.03}^{+0.03}$ & $0.36_{-0.06}^{+0.08}$ & $127_{-12}^{+17}$ & $2.6_{-0.3}^{+0.5}$ & $8.5_{-1.8}^{+6.1}$  & $-1.8_{-0.6}^{+0.5}$ & $-1.9_{-0.5}^{+0.4}$ & $0.33_{-0.23}^{+0.22}$ & $-1.3_{-0.4}^{+0.2}$ \\
D3-13 & $\geq 52.3$ & $52.3_{-0.2}^{+0.1}$ & $52.9_{-0.2}^{+0.2}$ & $0.20_{-0.02}^{+0.01}$ & $0.43_{-0.06}^{+0.04}$ & $123_{-7}^{+6}$ & $2.4_{-0.1}^{+0.2}$ & $6.6_{-0.9}^{+1.9}$  & $-1.7_{-0.3}^{+0.2}$ & $-1.8_{-0.2}^{+0.2}$ & $0.31_{-0.12}^{+0.16}$ & $-1.3_{-0.2}^{+0.2}$ \\
\bf{14ae} & $\approx 50.2$ & $50.2_{-0.0}^{+0.0}$ & $50.5_{-0.0}^{+0.0}$ & $0.51_{-0.03}^{+0.03}$ & $0.40_{-0.03}^{+0.03}$ & $18_{-1}^{+2}$ & $2.6_{-0.1}^{+0.2}$ & $6.0_{-0.5}^{+0.4}$  & $-3.3_{-0.2}^{+0.2}$ & $-3.4_{-0.1}^{+0.2}$ & $0.48_{-0.09}^{+0.07}$ & $-3.7_{-0.0}^{+0.0}$ \\
14li\footnotemark[1] & $\approx 50.8$ & $51.1_{-0.2}^{+0.2}$ & $51.6_{-0.3}^{+0.3}$ & $0.35_{-0.05}^{+0.07}$ & $0.24_{-0.06}^{+0.07}$ & $40_{-8}^{+9}$ & $4.4_{-1.3}^{+2.0}$ & $23.6_{-11.8}^{+27.2}$  & $-1.7_{-0.8}^{+0.6}$ & $-1.9_{-0.7}^{+0.6}$ & $0.46_{-0.15}^{+0.12}$ & $-2.7_{-0.3}^{+0.3}$ \\
\bf{16fnl} & $49.3_{-0.1}^{+0.1}$ & $50.1_{-0.0}^{+0.0}$ & $50.4_{-0.1}^{+0.1}$ & $0.51_{-0.04}^{+0.03}$ & $0.35_{-0.04}^{+0.02}$ & $20_{-1}^{+1}$ & $2.7_{-0.1}^{+0.3}$ & $12.9_{-2.9}^{+5.0}$  & $-3.3_{-0.0}^{+0.1}$ & $-3.4_{-0.0}^{+0.1}$ & $0.30_{-0.04}^{+0.04}$ & $-3.8_{-0.1}^{+0.1}$ \\
&& $(49.5_{-0.0}^{+0.0})$\footnotemark[2] & $(49.7_{-0.0}^{+0.0})$\footnotemark[2] &&&&&&&&\\
\bf{15oi} & $\approx 50.8$ & $50.4_{-0.2}^{+0.2}$ & $50.9_{-0.3}^{+0.2}$ & $0.37_{-0.05}^{+0.09}$ & $0.29_{-0.06}^{+0.09}$ & $31_{-4}^{+5}$ & $3.8_{-1.2}^{+1.9}$ & $34.4_{-22.5}^{+23.2}$  & $-3.0_{-0.3}^{+0.8}$ & $-3.0_{-0.3}^{+0.8}$ & $0.19_{-0.06}^{+0.07}$ & $-3.4_{-0.3}^{+0.2}$ \\
16axa & $\approx 50.7$ & $50.3_{-0.1}^{+0.2}$ & $50.4_{-0.1}^{+0.2}$ & $0.78_{-0.06}^{+0.05}$ & $0.27_{-0.03}^{+0.04}$ & $29_{-4}^{+6}$ & $3.9_{-0.7}^{+1.1}$ & $31.1_{-7.0}^{+18.6}$  & $-2.8_{-0.4}^{+0.3}$ & $-2.8_{-0.4}^{+0.3}$ & $0.10_{-0.03}^{+0.05}$ & $-3.9_{-0.1}^{+0.2}$ \\
\bf{11af} & $50.6_{-0.0}^{+0.0}$ & $50.7_{-0.0}^{+0.0}$ & $50.9_{-0.1}^{+0.1}$ & $0.62_{-0.15}^{+0.05}$ & $0.35_{-0.12}^{+0.04}$ & $29_{-2}^{+2}$ & $2.8_{-0.3}^{+2.3}$ & $14.1_{-4.5}^{+25.7}$  & $-2.9_{-0.4}^{+1.0}$ & $-3.0_{-0.4}^{+1.0}$ & $0.26_{-0.05}^{+0.05}$ & $-3.4_{-0.1}^{+0.1}$ \\
\hline
09ge & $52.0_{-0.6}^{+0.1}$ & $50.5_{-0.0}^{+0.0}$ & $50.6_{-0.0}^{+0.0}$ & $0.64_{-0.02}^{+0.02}$ & $0.27_{-0.01}^{+0.02}$ & $36_{-1}^{+1}$ & $3.9_{-0.4}^{+0.3}$ & $30.8_{-3.2}^{+2.3}$  & $-2.6_{-0.3}^{+0.2}$ & $-2.6_{-0.3}^{+0.2}$ & $0.09_{-0.01}^{+0.02}$ & $-3.6_{-0.0}^{+0.0}$ \\
09djl & -- & $51.0_{-0.2}^{+0.3}$ & $51.2_{-0.3}^{+0.4}$ & $0.58_{-0.12}^{+0.13}$ & $0.21_{-0.06}^{+0.07}$ & $26_{-3}^{+4}$ & $5.1_{-1.2}^{+2.3}$ & $23.5_{-11.2}^{+38.3}$  & $-1.6_{-0.4}^{+0.4}$ & $-1.7_{-0.4}^{+0.4}$ & $0.43_{-0.18}^{+0.21}$ & $-3.1_{-0.3}^{+0.4}$ \\
TDE2 & -- & $51.5_{-0.1}^{+0.1}$ & $52.1_{-0.1}^{+0.1}$ & $0.25_{-0.03}^{+0.03}$ & $0.09_{-0.05}^{+0.08}$ & $40_{-13}^{+36}$ & $11.0_{-5.2}^{+12.1}$ & $93.1_{-43.1}^{+107.6}$  & $-1.1_{-0.5}^{+0.4}$ & $-1.7_{-0.5}^{+0.7}$ & $0.86_{-0.33}^{+0.09}$ & $-2.1_{-0.1}^{+0.1}$ \\
TDE1 & -- & $51.0_{-0.2}^{+0.2}$ & $51.3_{-0.2}^{+0.3}$ & $0.45_{-0.11}^{+0.10}$ & $0.24_{-0.04}^{+0.05}$ & $52_{-17}^{+23}$ & $4.6_{-1.2}^{+1.4}$ & $38.5_{-15.1}^{+22.2}$  & $-1.9_{-0.5}^{+0.7}$ & $-1.9_{-0.5}^{+0.6}$ & $0.19_{-0.11}^{+0.22}$ & $-2.9_{-0.2}^{+0.3}$ \\
16aaa & $\approx 52.7$ & $51.2_{-0.4}^{+0.6}$ & $51.3_{-0.4}^{+0.6}$ & $0.88_{-0.09}^{+0.06}$ & $0.23_{-0.09}^{+0.09}$ & $33_{-5}^{+10}$ & $4.6_{-1.7}^{+2.9}$ & $21.4_{-11.0}^{+40.5}$  & $-1.6_{-0.6}^{+0.7}$ & $-1.7_{-0.6}^{+0.5}$ & $0.37_{-0.26}^{+0.30}$ & $-3.0_{-0.4}^{+0.6}$ \\
\hline
\end{tabular}
\end{center}
\footnotetext[1] {this event had significant additional X-ray emission contemporaneous with the optical and UV emission, however we only include an analysis of the energy emitted in optical and UV wavelengths in this work. The energy and efficiency estimates for this event are lower limits.}
\footnotetext[2]{values in parentheses were calculated from fits without accounting for host extinction. iPTF16fnl preferred fits with high host extinction, and the literature calculations did not include host extinction (see Section~\ref{sec:calorimetry} for more information).}

\caption{
The transients in the table are organized as follows: The first 9 events have UV detections during the same time period as the optical detections. The events in bold have observations at or before the light curve peak. Systematic errors for parameters are listed at the top of their respective columns, throughout the text we include the systematic errors in the errors quoted for the parameters. Systematic errors for $t_{\rm peak}$ and efficiency ($\epsilon$) are taken from \citet{Mockler:2019a}, additional errors were calculated using the method described in the same paper and are based on the uncertainty in the stellar mass-radius relation. 
{\bf Column descriptions:} (1) transient names; (2) Bolometric energy estimates from the literature. The methods used to calculate these estimates are described below in this caption; (3) Bolometric energy estimates from the \mosfit fits, integrated over the same time period used for the literature energy estimates in column 2. The literature energy estimate for PTF09ge is from late-time dust emission and therefore the \mosfit energy estimate for column 3 for this event was integrated over the time period of the initial optical observations presented in \citet{Arcavi:2014a}; (4) Column 2 divided by column 3; (5) The peak bolometric luminosity multiplied by the timescale from first fallback to peak luminosity, divided by column 3; (6) The $\Delta t$ between when the first stellar debris falls back to pericenter (`first fallback') and the time of peak luminosity. This is necessarily less than the time from disruption to peak; (7) \& (8) $\rm t_{\rm 50}$ and $\rm t_{\rm 90}$ are the respective times when 50\% and 90\% of the total energy is radiated. The first 5\% and last 5\% are excluded from the integral. In columns 7 \& 8 they are scaled by the peak timescale calculated from first fallback; (9) The average observed efficiency, defined as $\rm E_{bol}/\Delta M c^2$, where $\Delta \rm M$ is the total amount of mass that is bound to the black hole; (10) The peak observed efficiency, defined as $\rm L_{peak}/\dot{M}_{peak} c^2$; (11) Eddington ratio at peak luminosity; (12) The minimum amount of mass required to generate the integrated energy if the conversion from mass to energy were 100\% efficient ($\rm E_{bol}= \Delta M_{min} c^2$).
{\bf Notes on literature energy estimates:} 
The value for PS1-10jh was calculated by integrating the light-curve model using a lower limit for the temperature and luminosity \citep{Gezari:2012a}. The values for both D1-9 and D3-13 were calculated by integrating a $t^{-5/3}$ power law starting at $t_{\rm discovery}$ (after $t_{\rm peak}$) using the lower limits to the blackbody temperature and luminosity \citep{Gezari:2008a}. The value for PTF09ge was calculated from IR dust emission, motivating that there is additional radiated energy not observed in the initial optical and UV light curve \citep{van-Velzen:2016b}.  The values for ASASSN-14ae, ASASSN-14li, ASASSN-15oi, iPTF16fnl, iPTF16axa, and OGLE16aaa were calculated by integrating the blackbody fits to the observed optical and UV light curves \citep{Holoien:2014a, Holoien:2016a, Holoien:2016b,Blagorodnova:2017a, Hung:2017a, Wyrzykowski:2017a}. According to \citet{Holoien:2016a}, the blackbody fit for ASASSN-14li was `dominated by systematic errors'. For PS1-11af, \citet{Chornock:2014a} calculated the radiated energy by using a constant bolometric correction to the light curve from a blackbody fit 10 rest-frame days after peak.}
\label{table:Lptp}
\end{table*}

In Table~\ref{table:Lptp}, we also compare our total bolometric energy estimates with those derived directly from fitting the observational data. First, we integrate our bolometric luminosity curve to only include the time ranges used in the literature calculations. We find that most of our calculations are similar to previous estimates (see columns 2 and 3 of Table~\ref{table:Lptp} for the energy estimates, and the caption for a list of literature sources). We discuss the measurements that are inconsistent with the literature below.

For the transient iPTF16fnl, we calculate a significantly higher total energy when integrating over the same time period as the observations (we find $\rm log_{10} E_{bol} = 50.1^{+0.1}_{-0.1}$, \citealt{Blagorodnova:2017a} find $\rm log_{10} E_{bol} = 49.3^{+0.1}_{-0.1}$). Our original fit preferred a relatively high value for the host extinction ($\rm log_{10} N_{H,\; host} = 20.9^{+0.4}_{-0.4}$, we include host extinction as a parameter in the mcmc fit of the light curve). The literature calculation for iPTF16fnl does fit for host extinction, but unlike this work found the best fit to prefer no host extinction. We fit iPTF16fnl a second time without correcting for any host extinction, and found a lower value for the energy radiated during observations ($\rm log_{10} E_{bol} = 49.5^{+0.1}_{-0.1}$), consistent with \citet{Blagorodnova:2017a}. However, this fit has a likelihood score that is about 10 points lower than the fit with host extinction (160 vs 173), therefore we use the fit that includes a host extinction correction in this work. We do include the energy estimates from the fit without a host extinction correction in Table~\ref{table:Lptp}, below the values for the fit that includes host extinction.

We expect most of the literature estimates of the total energy emitted by transients in our catalog to be less than the values we derive for the total energy (column 4 of Table~\ref{table:Lptp}). This is because we integrate out to much later times than is possible with existing observations, and we always include the peak of the transient even when the event was caught in the decay phase. There are, however, two events that have literature estimates of the total energy that are significantly higher than those presented in Table~\ref{table:Lptp} (column 4). PTF09ge and OGLE16aaa both have literature estimates of the total bolometric energy that are higher than what we derive. In the case of PTF09ge, the literature value is calculated using IR emission measurements taken after the initial optical/UV light curve \citep{van-Velzen:2016a}. The fact that this method finds a higher value for the energy might imply that there is additional emission that was not captured by the initial optical and UV observations but was effectively reprocessed by dust. For OGLE16aaa, it is unclear why our calculation of the energy is lower than the one presented in the literature. According to \citet{Wyrzykowski:2017a}, the literature estimate is derived using blackbody fits to the optical and UV light curve. While our original fit was only to the optical light curve, we re-fitted the event to also include the sparse UV data and obtained a similar value for the total emitted energy, with a slightly reduced statistical error ($\rm log_{10}E_{bol} = 51.4^{+0.3}_{-0.2}$ erg). 

One of the liveliest debated issues in the TDE field concerns the nature of the luminosity decay after peak. This is in part because the fallback rates do not converge to the canonical $t^{-5/3}$ power law until well after peak, if they converge to it at all \citep[e.g.][]{ Guillochon:2013a}. The $t^{-5/3}$ power law decay rate is derived by assuming that the derivative of the mass-energy distribution ($dM/dE$) is constant and the orbit can be approximated as parabolic. This is a reasonable approximation for the material near the core of the star (the material least bound to the black hole) in a full disruption. However, as numerous papers have highlighted, this is never a good approximation at early times, and for many disruptions it is also not a good approximation at late times \citep[e.g.,][]{Gafton:2019}. Other power law decay rates have been predicted for the late-time emission from partial disruptions ($t^{-9/4}$, see \citealt{Coughlin:2019b}), and for disk emission ($t^{-5/12}$, see \citealt{Lodato:2011a, Auchettl:2017a}). These power laws do a better job of approximating the emission decay rates for the relevant flares at the relevant epochs, however, like the $t^{-5/3}$ power law, they have limited applicability. For this reason, we caution against approximating the entire decay rate of the emission as a single power law when calculating the total radiated energy. At the very least, the late-time asymptotic behavior must be treated separately from the early time emission near the peak of the light curve  \citep{van-Velzen:2019a}.

From the evolution of the cumulative energy, we also derived $t_{90}$ ($t_{50}$) values for all events, which we define here as the time frame during which $90\%$ ($50\%$) of the bolometric energy is accumulated, not including the first and last $5\%$. We clarify that we do not define $t_{50}$ as the time frame when the middle 50\% of the energy is emitted, as it is often defined for gamma-ray bursts, rather we define it as the time when the first 50\% of the energy has been radiated (excluding the first and last 5\%). Calculating these estimates required using the \mosfit model to extrapolate the bolometric luminosity out past the observations. In Table~\ref{table:Lptp} we list the values for $t_{50}/t_{\rm peak}$ and $t_{90}/t_{\rm peak}$ for each fitted event in Table~\ref{table:Lptp}.  We find that $t_{50}$ can occur anywhere from a month to more than a year after peak, while reaching $t_{90}$ takes significantly more than a year and up to $\approx 10$ years in some cases. This means that while observations of TDEs often extend out to $t_{50}$, they rarely extend out to $t_{90}$, contributing to the underestimation of the total energy radiated that we discussed above.

The compilation of the radiated energy inventory shown in Table~\ref{table:Lptp} provides a robust lower limit on the integrated effects of the energy transfer involved in TDEs. The total observed isotropic energy (column 3 in Table~\ref{table:Lptp}) that is radiated at optical and UV wavelengths is usually larger than that emitted in X-rays during the main flare. This is predominantly due to the fact that $\nu F_\nu$ typically peaks closer to the UV than to the X-rays (at least near the peak of the light curve). However, we note that this is not always the case and it still possible that previous X-ray observations could have missed the time when the X-ray luminosity was highest, resulting in a underestimate for some of the events \citep{Auchettl:2017a}. It is possible that we are missing X-ray emission due to inclination effects if our observational line of sight is edge on through the accretion disk, as opposed to closer to the poles (where jet emission would be beamed, \citealt{Dai:2018}). There are also recent reports of an excess of late-time emission for some TDEs \citep{van-Velzen:2019a} and we discuss the contribution of this emission to the total energy inventory in Section~\ref{sec:late-time}.

\section{Efficiency}\label{sec:efficiency}
Estimating the radiative efficiency is appropriate for our purpose of uncovering the energy transfer mechanisms  in tidal disruption flares.
The total efficiency ($\epsilon$), as defined by $E_{\rm bol} = \epsilon M_{\rm bound} c^2$, parameterizes the conversion from rest mass to electromagnetic energy  and provides a robust limit on the radiative efficiency. Here $M_{\rm bound}$ is the total mass bound to the black hole after the disruption, which is $\approx M_\ast/2$ for a full disruption.

As discussed in Section~\ref{sec:calorimetry}, we are able to provide a robust estimate of the total energy radiated during these events. However, more complicated modelling is required to estimate the efficiency of conversion from rest mass to electromagnetic energy. To calculate the radiative efficiency we first need a model that accurately estimates  the mass supply into the vicinity of the black hole. As described in the previous section, we use the results from hydrodynamic simulations in order to provide an estimate of the rate of mass return to pericenter. The bolometric luminosity can then be simply written as $L_{\rm bol}=\epsilon\dot{M}c^2$. However, the magnitude of the mass fallback rate ($\dot{M}$) is also dependent on the mass ($M_{\rm h}$) and spin ($a$) of the black hole, the mass of the star ($M_\star$), and the impact parameter of the disruption ($\beta$). Luckily, the mass of the black hole  significantly changes the peak timescale \citep{Mockler:2019a} while the impact parameter noticeably alters the shape of the fallback curve \citep{Guillochon:2013a, Gafton:2019}. As a result, the effects caused by changing these two parameters can generally be disentangled from the effects caused by altering the efficiency when fitting the light curves. It is, however, more difficult to break the degeneracy between the mass of the star and the efficiency, which leads to large systematic uncertainties in the determination of these parameters \citep{Mockler:2019a}.

As we mentioned briefly in Section~\ref{sec:calorimetry}, we have re-run the fits presented in \citep{Mockler:2019a} with the efficiency prior range expanded by an order of magnitude, from  $5\times 10^{-3} - 4 \times 10^{-1}$ to  $5\times 10^{-4} - 4 \times 10^{-1}$. As the fits are to the same observational data as in \citep{Mockler:2019a}, the light curves we obtain are very similar. With this added flexibility, many of the events prefer lower efficiencies and higher stellar masses than those presented in \citet{Mockler:2019a}.

However, we caution that the degeneracy between the efficiency and the stellar mass makes it difficult to pinpoint a particular stellar mass, and therefore it is not clear that these lower efficiency, higher stellar mass fits are more accurate than the previous fits run with a smaller efficiency prior range. For example, we note that in these new fits, two of the events with the lowest radiative efficiencies prefer very high stellar masses that are strongly disfavored by IMF functions (PS1-10jh prefers $M_*=8.2^{+29.7}_{-6.4} M_{\odot}$ and  ASASSN-14ae prefers $M_*=5.4^{+22.0}_{-4.3} M_{\odot}$). 
Although there is significant uncertainty in the determination of the mass of the disrupted star and the associated radiative efficiency, we note that the degeneracy between them does not noticeably change the determination of the bolometric energy measured, as clearly shown in Figure~\ref{fig:eff_mstar}.  Additionally, the black hole masses remain consistent with previous estimates, as do the scaled impact parameters for all fits except iPTF16axa, which now prefers a full disruption rather than a partial disruption. iPTF16axa does not have measurements near peak or strong upper limits, and therefore the rise and the peak of the light curve are not well constrained, which is likely why both a full disruption and partial disruption can fit the light curve. 

\subsection{The mass distribution of tidally disrupted stars}

\begin{figure}[t]\centering
\includegraphics[width=1 \linewidth,clip=true]{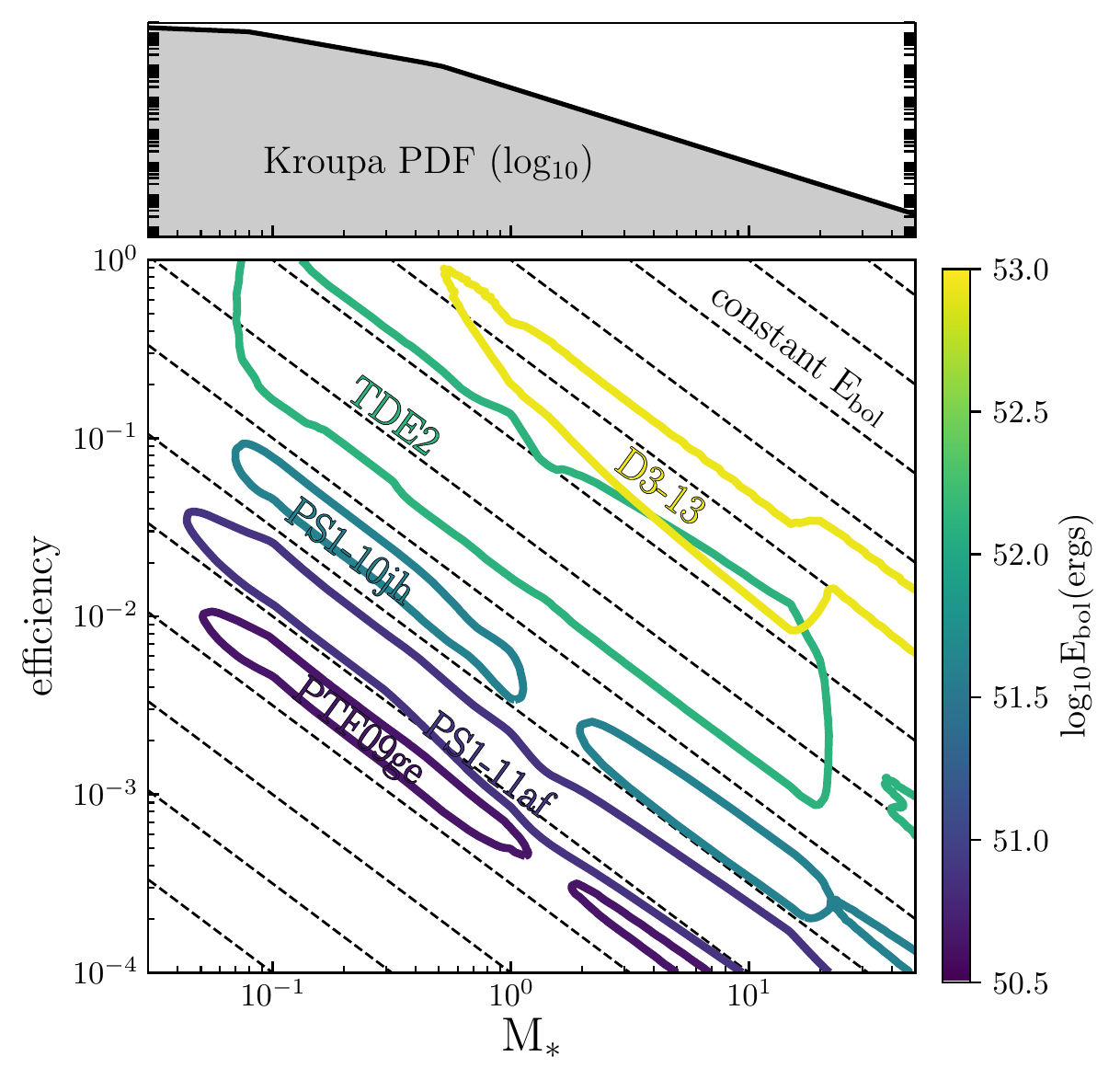}
\caption{The role of stellar mass in TDEs. {\it Bottom Panel}: The likelihood contours for stellar mass and efficiency parameters derived for five TDEs with a spread of bolometric energy values. The contours have been calculated by taking the parameter values from the converged walker distributions from the \mosfit fits, varying the efficiency parameter between $0.0001 - 1$ and the stellar mass between $0.01-100$, and recalculating the likelihood of each parameter combination. We kept the values from the converged walker distributions constant for all parameters except for $M_{\ast}$ and efficiency. The contours were arbitrarily chosen as $\rm log_{10}likelihood_{max} - 30$ as this contour value clearly shows the stellar mass - efficiency degeneracy for all plotted fits. 
Also shown are dashed lines denoting constant bolometric energy. The lines show constant $\rm E_{bol}$ for a given $\rm M_{accreted}/M_{\ast}$. Impact parameter (and therefore $\rm M_{accreted}/M_{\ast}$) vary from TDE to TDE, therefore the value of these lines can vary from event to event. However, the general trend of increasing $\rm E_{bol}$ as one moves up and to the right in the plot is clearly shown in the data. {\it Top Panel}: The Kroupa IMF is plotted for comparison \citep{Kroupa:2001a}.}
\label{fig:eff_mstar}
\end{figure}

We probe the degeneracy between stellar mass and radiative efficiency by taking the best fit walker distributions derived in \citet{Mockler:2019a}, and varying the efficiency and stellar mass parameters over the following ranges: $0.0001<1$; $0.01M_\odot < M_\star < 100M_\odot$. This makes it possible to visualize a large region of likelihood space for these parameters, and the resulting likelihood distributions for a subset of the transients are shown in Figure~\ref{fig:eff_mstar}. For the plot we chose transients that covered a wide range in parameter space, and have limited the number plotted because adding additional events would overlap. For a given total radiated energy, we find that there is a clear degeneracy between stellar mass and efficiency. However, the spread in total radiated energy is, as expected, small. 

Most common theories of tidal disruption rates foresee the
disrupted stars coming from within the radius of influence of the SMBH \citep[e.g.,][]{Frank:1976a}.
Each star within this region is expected to trace out a complicated orbit under the combined  influence of all other stars and the SMBH
itself. In these crowded regions, two-body interactions alter the distribution of stars on long timescales and, as a result, we expect the disrupted stars to come from near the peak of the IMF (see top panel in Figure~\ref{fig:eff_mstar}). If this is the case, we expect the radiative efficiencies to be $\gtrsim 10^{-2}$.

Recently, there have been several theories put forward to explain the surprising finding that tidal disruption events preferentially occur in an unusual sub-type of galaxies known as E+A galaxies \citep{Arcavi:2014a, French:2016a, Law-Smith:2017b}. These theories include an overdensity of stars near the SMBH \citep{Stone:2018a, Law-Smith:2017b}, the presence of a SMBH companion \citep{Arcavi:2014a, Li:2015a}, and star formation in eccentric disks around SMBHs \citep{Madigan:2018}. Common to all of these theories is the idea that stars deep in the potential well of the SMBH interact with one another coherently, resulting in rapid angular momentum evolution \citep{Rauch:1996a}.
Within this formalism, higher mass stars could, in principle, be preferentially disrupted (compared to the IMF), and as such, systematically lower radiative efficiencies might be possible (Figure~\ref{fig:eff_mstar}). 

\subsection{The efficiency of super-Eddington accretion}
Radiative efficiency can be temporarily reduced near the peak of the flare if the mass accretion rate  exceeds the Eddington limit of the SMBH. In this case much of the returning debris must either escape in a radiatively driven wind or be accreted inefficiently \citep{Ramirez-Ruiz:2009a, Strubbe:2009a}. In the second case, a sizable fraction of the mass would be fed to the black hole far more rapidly than it could be accepted if the radiative efficiency were high. This can add to the difficulty of measuring the efficiency of conversion from the rest mass energy to luminosity. For example, in \citet{Dai:2018}, the authors ran a state of the art simulation of a super-Eddington accretion disk around a massive black hole with a spin of 0.8 and a mass fallback rate of $\approx 15 \dot{M}_{\rm Edd}$ and found a reduced mass to luminosity efficiency of 2.7\%. According to the same model, events with significant optical emission required the observer to be viewing the event along the direction of the disk to allow for sufficient reprocessing of emission. As such, their mass to observed luminosity efficiency was even lower than 2.7\% (the majority of emission was beamed along the polar direction). This is much lower than the expected Novikov-Thorne efficiency of 12.2\% for a black hole with a spin of 0.8 \citep{Novikov:1973}. 

The \mosfit TDE model approximates the effect of the super-Eddington regime on the observed luminosity with a soft cut on the luminosity as it approaches $L_{\rm Edd}$, reducing the efficiency of conversion between $\dot{M}$ and $L_{\rm bol}$ near $L_{\rm Edd}$ \citep{Mockler:2019a}. The average efficiency and the efficiency at peak derived from our model are listed in Table~\ref{table:Lptp}, column 9. As expected, the peak efficiency is generally lower than the average efficiency, most noticeably for flares with peak luminosities near the Eddington limit. We note that using the peak efficiency to estimate either the total radiative efficiency or the total energy radiated during these events will lead to an underestimation in these quantities.   

\begin{figure}[t]
\centering
\includegraphics[width=1 \linewidth,clip=true]{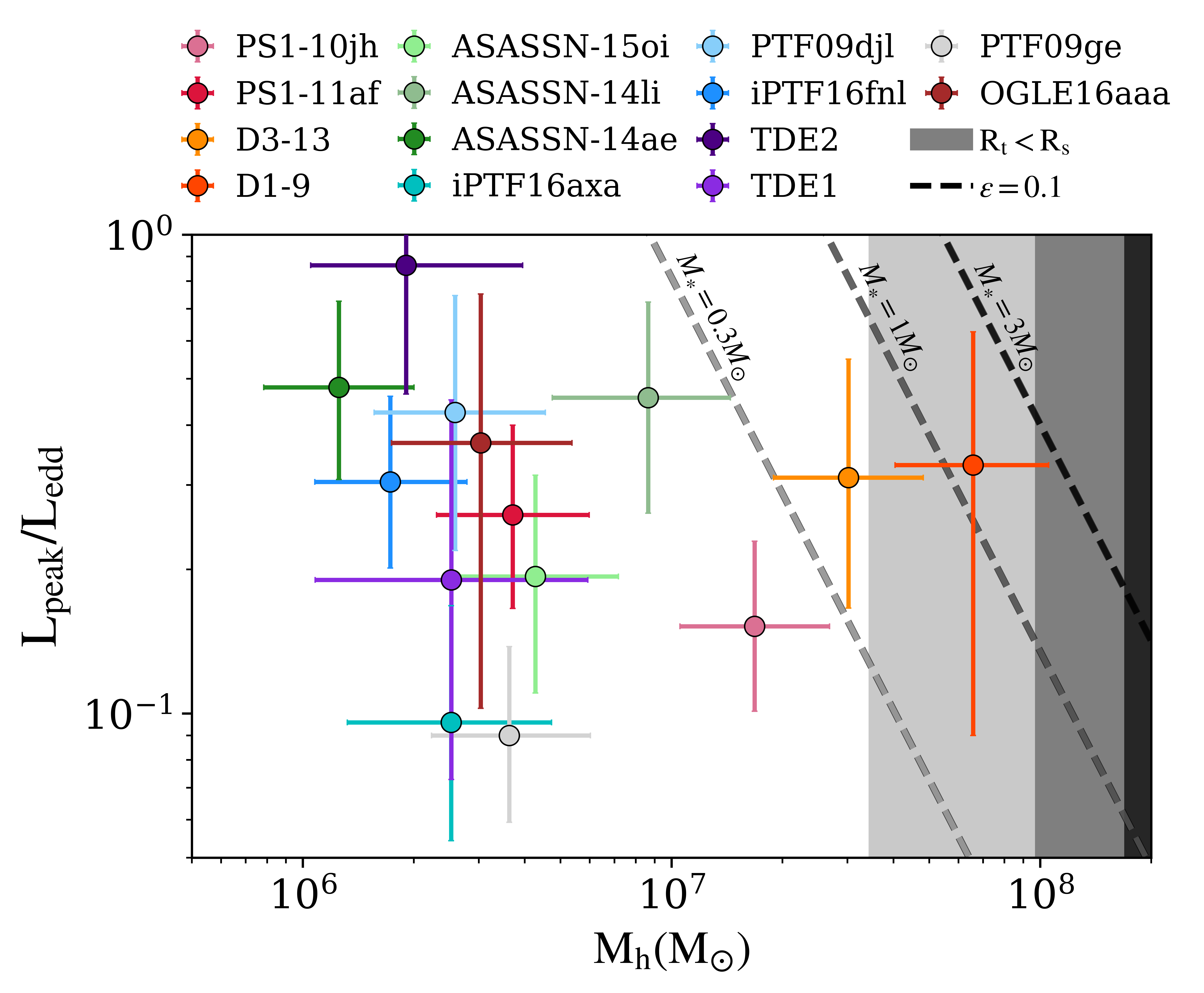}
\caption{
Using dashed lines, we plot the Eddington ratio for simulated flares using several stellar masses and assuming an efficiency of 0.1 at peak (a typical AGN efficiency that would produce super-Eddington flares for most TDEs). The shaded regions show were $\rm R_t<R_s$ -- the approximate point where stars will be disrupted inside the black hole's event horizon (the lightest shading corresponds to $0.3 M_{\odot}$, the medium shading to $1 M_{\odot}$, and the darkest shading to $3 M_{\odot}$). Eddington limited accretion has a maximum efficiency determined by the accretion radius, mass accretion rate, and Eddington limit, and the median value of the Eddington ratio for this sample is $\rm L_{peak}/L_{edd} = 0.3^{+0.3}_{-0.2}$. 
}
\label{fig:LpLedd_Mh}
\end{figure}

The maximum possible peak efficiency of a tidal disruption event is dependent on the Eddington limit of the SMBH, however it is also dependent on the mass of the disrupted star and the impact parameter of the disruption. In Figure~\ref{fig:LpLedd_Mh}, the dashed lines show $\rm L_{peak}/L_{edd}$ for full disruptions of different stellar masses, assuming a peak efficiency of 10\%. 

For example, black holes below $\approx 10^7 M_{\odot}$ will be super-Eddington at 10\% peak efficiency for (full) disruptions of stars above $\approx 0.3 M_{\odot}$. This implies that most of the black holes in our sample can only reach peak efficiencies of 10\% if they disrupt low mass stars, or they break the Eddington limit. On the other hand, as the mass of the black hole increases past $10^7 M_{\odot}$, the Schwarzschild radius begins to encroach on the tidal radius for main-sequence stars. Once the Schwarzschild radius grows larger than the tidal radius, most disruptions will occur within the black hole's event horizon, and there will be no observable flares.\footnote{The exact value of the maximum black hole mass able to produce an observable flare depends on the spin of the black hole and the stellar structure of the star. It also depends on whether or not partial disruptions are included in the calculation, as stars can be partially disrupted at larger radii than are required for full disruptions. This calculation must be done using general relativity, as the tidal forces at a given radius are noticeably different in GR for high mass black holes \citep{Servin:2017}}

In Figure~\ref{fig:LpLedd_Mh} these disallowed regions of parameter space are shaded in gray, providing additional constraints on the stellar masses for the tidal disruption flares around the largest black holes in this sample. The most massive black hole in this sample, the black hole  in the host galaxy of D1-9 (J022517.0-043258), can only disrupt stars $\gtrsim 0.6 M_{\odot}$.  The total energy released in the flare around this black hole is also quite large, and requires the conversion of $0.5 M_{\odot}$ of mass to energy if the efficiency is $\sim 10\%$, and a correspondingly larger amount of mass for lower efficiencies (from Table~\ref{table:Lptp}, $\rm 10^{53} ergs \approx 0.1 \times 0.5 M_{\odot} \times c^2$). Therefore, both the mass of the host black hole and the luminosity of the flare suggest that the disrupted star  for the TDE D1-9 was above the peak of the IMF, and likely $\gtrsim 1 M_{\odot}$.

While there is scatter in the Eddington fraction for this sample of TDEs, with values varying between $\sim (0.1 - 0.9) \; \rm L_{peak}/L_{edd}$, the median of the distribution is $0.3^{+0.3}_{-0.2} \;  \rm L_{peak}/L_{edd}$ (the median value calculated from the fits in \citet{Mockler:2019a} was $0.3^{+0.4}_{-0.2}$). This is similar to the median Eddington fraction derived for AGN \citep{Kollmeier:2006}.
However, if observed TDE rates are driven by the brightest events, we might expect most of the flares to be pushing up against their black holes' respective Eddington limits \citep{Kochanek:2016a}. 

Very high Eddington ratios would likely produce flattened light curves that would no longer appear to follow the shape of the mass fallback rates near the peak of the light curve (or wherever the mass fallback rate is super-Eddington), due to the significant mass wasted during super-Eddington accretion. Most of the events in this sample (and all of those with well-sampled observations at peak) appear to follow the shape of the expected mass fallback rates and do not push up against the Eddington limit. We note that the fit for TDE2 does prefer a particularly high Eddington fraction ($\approx 0.9$) and its light curve is flatter than the rest of the sample, as shown in Figure 1 of \citet{Mockler:2019a}. Unfortunately, as there are no observations at peak for TDE2, that section of the light curve is extrapolated and has large uncertainties, making it is difficult to determine the accuracy of the flattened portion of the light curve.

\section{Discussion}\label{sec:disc}

\subsection{Late-time energy release}\label{sec:late-time}
As we have shown in \citet{Mockler:2019a}, the luminosity evolution of tidal disruption events broadly follows the fallback rate of debris. As shown in column 8 of Table~\ref{table:Lptp}, $t_{\rm 50}$ of the radiated energy occurs long after $t_{\rm peak}$ for the majority of the fitted events ($t_{\rm 50}$ occurs later than 100 days after peak for 7 of the 14 events, and $t_{\rm 50} > 3 \times t_{\rm peak}$ for 9 of the 14 events). As a result, a significant fraction of the total energy is available to be emitted at late times. To check our estimates of the total bolometric energy released in these events, we need observations out to at least $t_{\rm 50}$. 

A recent search by \citet{van-Velzen:2019a} for late-time emission from TDEs at times $\gtrsim 10\times t_{\rm peak}$ found that 10 transients were still emitting at FUV wavelengths (the optical emission had faded below host levels). The same study found that power law fits to early-time FUV emission under-predicted the late-time FUV emission for PTF09ge, PTF09djl, ASASSN-14ae, iPTF16fnl, PS1-10jh, and ASASSN-14li, implying that at these frequencies and at times $\gtrsim 10\times t_{\rm peak}$, the emission was decaying at a slower rate than predicted by the extrapolations of power law fits to observations near peak.

\begin{figure}[t]
    \begin{center}
    \includegraphics[width=1 \linewidth,clip=true]{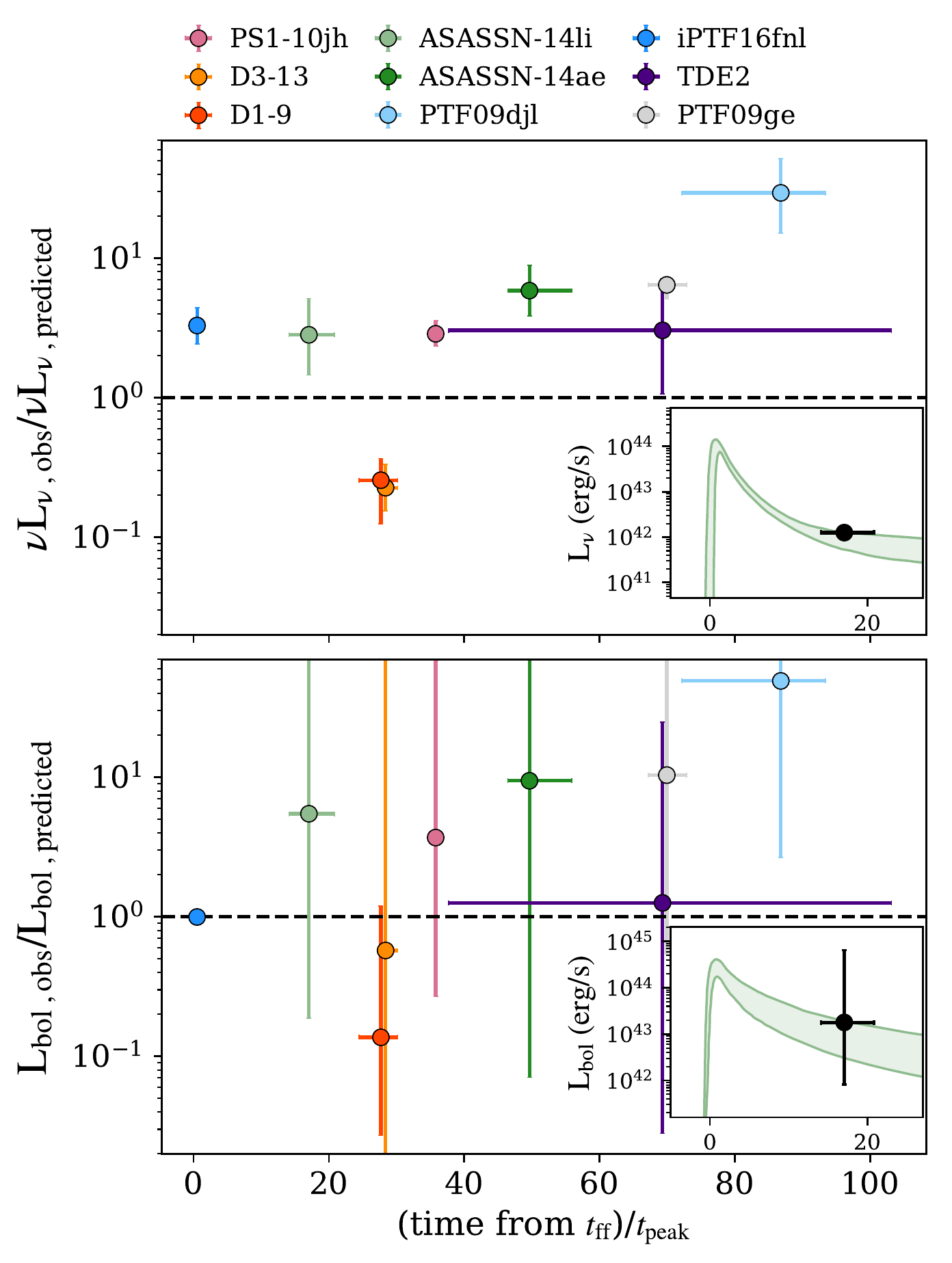}
    \end{center}
    
    \caption{We compare the predicted values for $\rm \nu L_{\nu}$ and $\rm L_{bol}$ from the \mosfit model to the observed FUV luminosities and corresponding $\rm L_{bol}$ estimates from \citet{van-Velzen:2019a}. The `predicted' values are calculated by extrapolating the \mosfit fits out to the relevant observation times. The x-axis is in units of time from first mass fallback normalized by the peak timescale (this removes the influence $\rm M_h$ has on the light curve timescale). The uncertainties in the model luminosities come from the fits to early-time data (this is easily seen for the inset light curves). The model uncertainties are then added in quadrature with the uncertainties from the observations.
    \textit{Top Panel:} $\rm \nu L_{\nu}$ is calculated at $ \gamma = 1500$\AA. $\rm \nu L_{\nu, \; observed}$ is derived from HST observations with the F125LP and F150LP filters. The \mosfit extrapolation set the minimum photosphere radius to $\rm R_{isco}$. 
    \textit{Bottom Panel:} 
    To calculate $\rm L_{bol, \; predicted}$, the model efficiency was assumed to be constant at late times. $\rm L_{bol, \; observed}$ was estimated from blackbody fits to observations in the F125LP and F150LP filters. There is significant uncertainty in the estimate of $\rm L_{bol, \; observed}$ due to the lack of SED coverage. We note that iPTF16fnl has a late-time observation in just one filter, therefore $\rm L_{bol, \; observed}$ was calculated using the temperature estimated from blackbody fits at early times and we are thus unable to accurately estimate  the associated uncertainties for this measurement.
    \textit{Inset Plots:} We plot $\rm \nu L_{\nu}$ and $\rm L_{bol}$ light curves for ASASSN-14li to give the reader an idea of how the discrepancies between the model and the observations shown in this plot map to a typical light curve. The green shaded curves are the model fits extrapolated out past the late time observations, the black points are the late time observations. The x-axes are the same as the larger plots, (time from $t_{\rm ff}$)/$t_{\rm peak}$.}
    \label{fig:Lobs_Lpred}
\end{figure}

To compare the observed late-time emission to predictions from the \mosfit model, we let $L_{\rm bol} = \epsilon \dot{M} c^2$, and used the photosphere evolution determined from our fits to the early light curve (with a minimum photosphere radius = $R_{\rm circ}$) to get estimates of the FUV emission at late times ($\gtrsim 10 \times t_{\rm peak}$). This assumption that the photosphere evolution can be extrapolated from early times is likely inaccurate at very late times ($\gtrsim 10 \times t_{\rm peak}$), however we use it to compare with the predictions made by extrapolating the power law fit to the early light curves. We find that our estimates of the late time $\nu L_{\nu}$ values sometimes under-predict and other times over-predict the observed values, as shown in Figure~\ref{fig:Lobs_Lpred}. The early time photosphere evolution no longer provides a good description for the late-time emission, and we find that like \citet{van-Velzen:2019a}, we also under-predict the FUV luminosity for PTF09ge, PTF09djl, iPTF16fnl, ASASSN-14ae, ASASSN-14li, and PS1-10jh. In general, the discrepancies between our predictions and observations are smaller than those calculated in \citet{van-Velzen:2019a}, and our predictions are within an order of magnitude of the observed values for all events except PTF09djl. 

In the bottom panel of Figure~\ref{fig:Lobs_Lpred} we also compare the predicted $L_{\rm bol}$ from our model to the value derived by \citet{van-Velzen:2019a} by fitting blackbody fits to the late-time observations. The values from our model come from assuming that the efficiency of conversion from mass to energy remains constant at late times, and $L_{\rm bol}$ continues to follow the mass fallback rate. The predicted values for $L_{\rm bol}$ are consistent with the blackbody fits within the quoted uncertainties for most events (the uncertainties are quite large as $L_{\rm bol, \; obs}$ is derived from observations in FUV bands only). There is enough energy in our 
\mosfit predictions to produce the observed FUV luminosity, and for the transients with measured FUV luminosities that are only a factor of a few different from the model predictions, it is likely that a different blackbody radius could reproduce the FUV luminosity. However, for the events with the largest discrepancies between the observations and the model predictions (e.g. PTF09djl), it is clear that a simple reprocessing model fails to provide a reasonable description of the observed FUV luminosity. 

One way to get better constraints on the energy released at late times is to obtain X-ray limits. Two of the events with late-time FUV measurements also have X-ray detections from \citet{Jonker:2020}. Both PTF09ge and ASASSN-14ae were detected by \textit{Chandra}, strengthening the case made by \citet{van-Velzen:2019a} that the bolometric luminosity of these events remains higher than expected from extrapolations of early time light curves (assuming the X-ray emission does not result from an underlying AGN). As we discuss in Section~\ref{sec:calorimetry}, there is further evidence that PTF09ge was more energetic than its initial UV/optical light curve suggests. The bolometric energy estimated from IR dust emission is significantly higher than the estimate from blackbody fits to the light curve (see Table~\ref{table:Lptp} and \citet{van-Velzen:2016b}), which suggests the presence of additional high energy emission missed in the initial observations.  

In  \citet{van-Velzen:2019a}, the authors point out that the late-time FUV observations from flares around larger mass black holes are closer to their predictions from early light curve extrapolations than the observations from smaller mass black holes. The smaller mass black holes required the addition of a second, shallower power law to match the data. However, the timescale of evolution of the light curve is strongly dependent on the black hole mass. Therefore, it is necessary to scale the light curve  by the black hole mass in order to effectively compare them. In fact, two of the most extreme under-predictions to observations are for the transients whose late time observations were taken after the largest number of peak timescales (PTF09djl and PTF09ge). Taken in this context, because the timescales of mass fallback are lengthened for larger mass black holes, the fact that \citet{van-Velzen:2019a} find more dramatic light curve evolution around smaller mass black holes is likely due to the fact that their mass fallback rate (and therefore their reprocessing layers and accretion rates) evolve on shorter timescales.

\subsection{Comparison to AGN}
While it is an important first step towards understanding how energy is generated in TDEs, this work leaves a number of important questions regarding the radiation physics of TDEs and exactly how and where the TDE emission is generated.

The simplest model for TDE emission is thermal radiation from an accretion disk at  $R_{\rm t}$, however it fails to accurately describe the observations. It predicts temperatures that are too high by a factor of about 5, and an optical luminosity more than an order of magnitude too dim \citep[e.g.,][]{Gezari:2008a}. This suggests that X-rays from accretion are being reprocessed into the optical part of the spectrum by a debris envelope with a much larger effective photosphere, perhaps 10-100 times $R_{\rm t}$. The origin of the extended gas remains unknown. It has been suggested that a large quasi-static envelope can be supported by radiation pressure \citep[e.g.,][]{Loeb:1997a}. An alternative explanation is that the extended gas instead results from strong accretion disk outflows \citep[e.g.,][]{Strubbe:2009a}.
Other works have suggested that the luminosity is not powered by accretion at all, but by shocks produced as material falls back, collides, and circularizes at large radii \citep[e.g.,][]{Piran:2015b}. In the absence of radiation transfer studies, it is difficult to distinguish between the proposed theories of the origin of TDE emission, and several basic questions about the spectrum formation in these environments remain open.
\begin{figure*}[t]
    \centering
    \includegraphics[width=   \textwidth,clip=true]{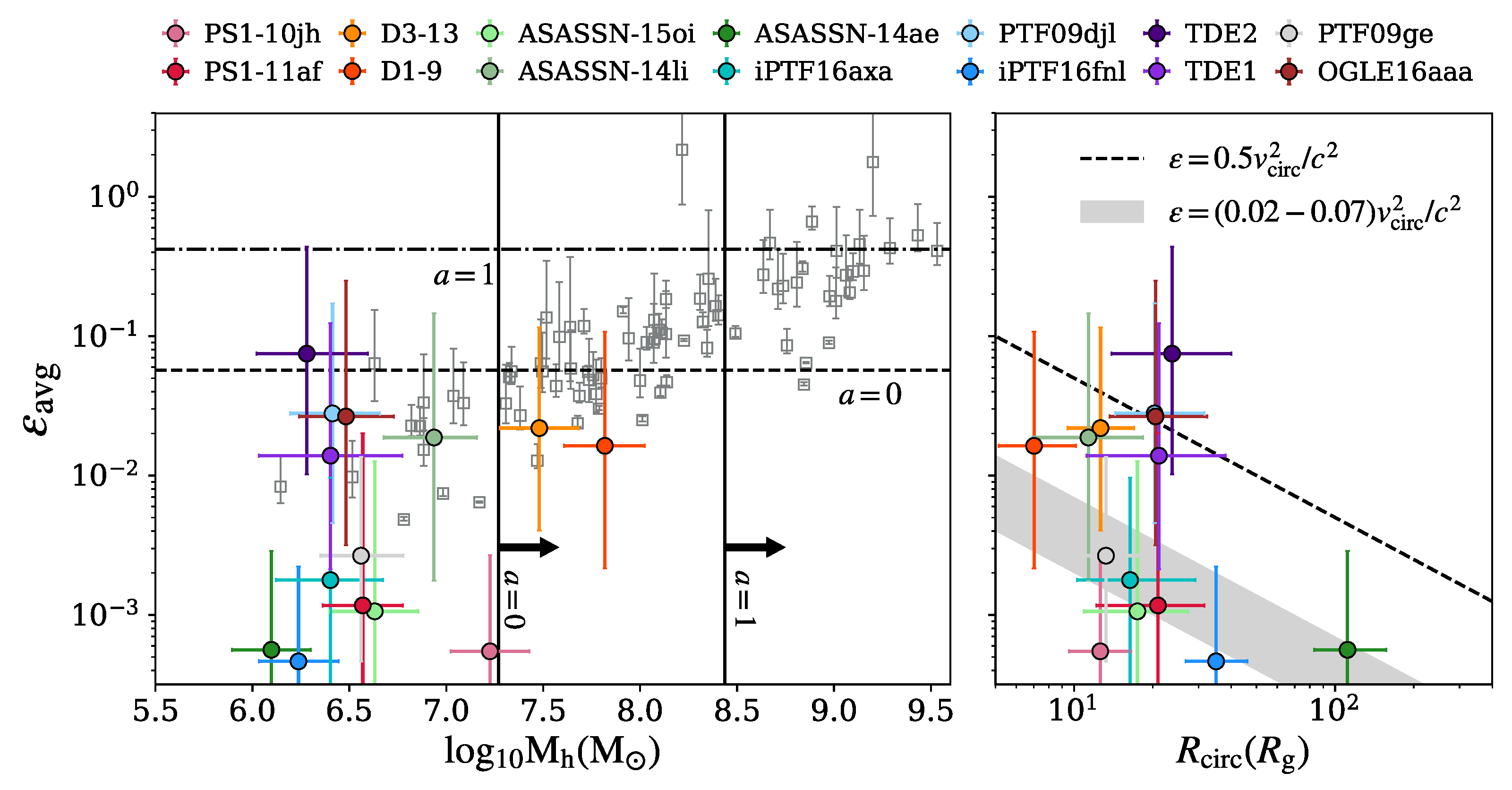}
    \caption{{\it Left Panel:} We compare the average efficiencies (integrated over the full light curve) of our sample of TDEs with AGN efficiencies from \citet{Davis:2011a} as a function of black hole mass. Our data is plotted as colored circles, while the AGN data from \citet{Davis:2011a} is plotted as gray squares. The trend of increasing efficiency with increasing black hole mass in the AGN data has been argued to be due at least in part to selection effects \citep{Raimundo:2011, Laor:2011}. We plot two vertical lines denoting where $\rm R_{t} < R_{isco}$ for a $1 M_{\odot}$ ZAMS star disrupted by black holes with spins of $a=0$ and $a=1$ respectively ($\rm R_{isco} = 3 R_{s}$ if $a=0$, $\rm R_{isco} = 0.5 R_{s}$ if $a=1$). {\it Right Panel:} We plot the average efficiencies of our sample versus the circularization radius in units of gravitational radii, with a dashed line denoting the maximum efficiency of conversion between kinetic energy (KE) and radiated energy at a given radius (assuming the gas is virialized post-collision). While the dashed line is the maximum theoretical efficiency of stream collisions at a given radius, simulations by \citet{Jiang-YF:2016} found the stream collision efficiency to be much lower -- the radiated energy was $\approx 2-7\%$ of the total KE. If these collisions occur at the circularization radius, we might expect the efficiencies to fall within the gray shaded region in the plot. However, the collision radius of the most bound debris will be much larger than the circularization radius unless the disruption is very deep and the black hole is very large, therefore the efficiency of the stream collisions will likely be much lower ($\rm R_{coll} \approx 6 \times R_{\rm circ}$ for the most bound debris in a full disruption of a solar mass star by a $5 \times 10^6 M_\odot$ Schwarzschild black hole) \citep{Jiang-YF:2016, Dai:2015a, Guillochon:2015b}.
    }
    \label{fig:DavisLaor}
\end{figure*}
As discussed in Section~\ref{sec:efficiency}, estimating the efficiency of conversion between $\dot{M}$ and $L_{\rm bol}$ helps us understand the nature of the processes producing the emission. The theoretically expected efficiencies for accretion disks around non-spinning and maximally spinning super-massive black holes are 0.057 and 0.42 respectively \citep{Novikov:1973}. Observational studies of AGN have found the mean accretion efficiency for AGN to be $\approx 0.08-0.1$ with large variation between individual AGN \citep{Marconi:2004a, Davis:2011a}. It is likely that the efficiency at peak for TDEs is lower than the average efficiency for AGN, simply because many TDEs approach their Eddington limit at peak (as discussed in Section~\ref{sec:efficiency}). However, we might expect the efficiency averaged over the full TDE evolution to be comparable to AGN if the majority of the luminosity is coming from accretion. 

In Figure~\ref{fig:DavisLaor} (left panel)  we compare the average efficiencies estimated by our model with the efficiencies measured for AGN in \citet{Davis:2011a}. There is significant spread in the measured efficiencies for the TDEs in our sample, and we find that while many of the efficiencies we estimate are consistent with AGN efficiencies within the large associated uncertainties, the lowest efficiencies in our sample are much lower than those expected from AGN. We also note that the radii of the innermost stable orbits ($\rm R_{isco}$) for the two most massive events in our sample are larger than the tidal radii for the disruptions of $1 M_\odot$ stars. For an accretion disk to form effectively, the disruption would likely need to occur outside of $\rm R_{isco}$, or a significant fraction of the debris will plunge directly into the black hole \citep{Gafton:2019}. In principal, this constraint can help narrow the range of possible stellar masses for TDEs, however in this case, D1-9 and D3-13 prefer masses $\gtrsim 1 M_\odot$ in both the fits with the expanded efficiency range used in this paper and the original fits in \citealt{Mockler:2019a}. We note that the trend of increasing efficiency with increasing AGN black hole mass in \citet{Davis:2011a} has been argued to be due in part to selection effects, however the plotted estimates are thought to be relatively accurate \citep{Laor:2011, Raimundo:2011}.

In Figure~\ref{fig:DavisLaor} (right panel) we plot the average efficiencies of our TDE sample versus the circularization radius in units of gravitational radius. We find that most of the efficiencies derived are also consistent within the errors to the maximum possible efficiencies from stream collisions, yet many of the events are in tension with the expected efficiencies from the stream collision model. Assuming gas remains approximately virialized, the maximum fraction of the rest mass energy available for dissipation (and therefore the maximum possible radiative efficiency) at a given radius from the black hole is $\approx 0.5 \times v_{\rm esc}^2/c^2 = 0.5 \times R_{\rm g}/R_{\rm conv}$, where $v_{\rm esc}$ is the escape speed at a given radius, $R_{\rm g}$ is the gravitational radius, and $R_{\rm conv}$ is the radius at which the rest mass energy is dissipated. This is a very conservative limit, and in reality the efficiency from stream collision is likely much lower (as we will discuss in the following paragraphs). Therefore, if most of the early time emission came from stream collisions or circularization processes instead of from accretion, these processes would need to occur close to the black hole for many of the transients in this sample ($\lesssim 100 R_{\rm g}$ for half of the events, see Figure~\ref{fig:DavisLaor}), and they would need to be very efficient at converting kinetic energy to radiation. 

A natural size scale for the accretion disk is the circularization radius ($R_{\rm circ}$). The circularization radius is calculated from the tidal radius and the impact parameter, $R_{\rm circ} = 2 R_{\rm p} = 2 R_{\rm t}/\beta $, and defines the radius of a circular orbit with the same angular momentum as an eccentric orbit with pericenter radius `$R_{\rm p}$'. The smallest possible radius at which stream collisions could occur is the pericenter radius, however the expected stream collision radius for the most bound debris is always larger than the circularization radius for full disruptions of main sequence stars by Scharwzschild black holes. The radius at which streams collide is dependent on the orbit of the bound debris, which can be approximated using the the mass and radius of the star, the mass (and spin) of the black hole, and the impact parameter of disruption \citep{Jiang-YF:2016, Dai:2015a}.

For example, given a Schwarzschild $5\times 10^6 M_{\odot}$ black hole, $\rm R_{coll} \approx 6  R_{circ}$ for a $0.3 M_{\odot}$ or $1 M_{\odot}$ star, and  $\rm R_{coll} \approx 17 R_{circ}$ for a $10 M_{\odot}$ star assuming critical disruptions of both stars ($\beta = 0.9$ for the  $0.3 M_{\odot}$ star,  $\beta = 1.8$ for the 1 and 10 $M_{\odot}$ stars; \citealt{Guillochon:2015b}). We see that $\rm R_{circ}$ provides an approximate size scale for the transition between where accretion processes occur and where stream collisions occur.

We thus conclude that it is only energetically feasible for stream collisions to power events with efficiencies close to the maximum efficiency at the circularization radius if collisions occurring near this radius have conversion efficiencies between kinetic energy and radiated energy that are larger than $50\%$. In \citet{Jiang-YF:2016}, the conversion rate between kinetic energy and radiated luminosity for stream collisions is found to be $\approx 2-7\%$ (or $\approx 4-14 \%$ of 1/2 of the total kinetic energy -- the fraction available to be radiated if the material is virialized), emphasizing that realistic efficiencies for stream collisions are likely much less than $50\%$.

About half of the events in this sample have efficiency estimates that are comparable to the maximum efficiency at the circularization radius, while the other half have efficiency estimates that are closer to 2-7\% of the KE at the circularization radius (of course there are still large systematic uncertainties, as discussed in Section~\ref{sec:efficiency}). We note that the degeneracy in $M_\ast -\epsilon$ does not significantly change this result, as using a higher/lower efficiency requires a lower/higher stellar mass and the circularization radius increases with stellar mass. Using a lower efficiency requires using a larger circularization radius, and vice versa, moving events diagonally in the $\epsilon - R_{\rm circ}$ plane. More precisely, $R_{\rm circ} \propto M_{\ast}^{1/3} R_\ast \propto M_\ast^{1/3 + \xi}$ -- the stellar radius, $R_\ast$, can be approximated as $R_\ast \propto M_\ast^\xi$ with $2/3\gtrsim \xi \gtrsim 1$ for most zero age main sequence (ZAMS) stars \citep{Tout:1996a}. Therefore, as you increase/decrease efficiency and decrease/increase $M_\ast$, you move along lines in $\epsilon - R_{\rm circ}$ space that are approximately parallel to $\epsilon \propto 1/R_{\rm circ} \propto v_{\rm circ}^2/c^2$. Of course, if the `scaled' impact parameter is constant (the approximate fraction of mass bound to the black hole remains the same), then the physical impact parameter ($\beta$) will change as one changes stellar structure. However, for main sequence stars it will generally not change the circularization radius by more than a factor of $\approx 2$, except in the case of super-critical disruptions ($R_{\rm circ} \propto 1/\beta$, and a full disruption of a lower mass ZAMS star occurs at $\beta \approx 0.9$, while a full disruption of a solar mass ZAMS star occurs at approximately $\beta \approx 1.8$). 

It remains possible that some fraction of TDEs are powered by stream collisions early on in their light curves, while others are powered by accretion. While it is likely that the flares with the highest measured efficiencies originate at small radii close to the black hole, and the flares with the lowest measured efficiencies originate further from the gravitational radius, it is clearly necessary to reduce the uncertainty in these efficiency estimates before we can derive more stringent conclusions. 

\subsection{Summary and Future Prospects}\label{sec:summary}
\begin{itemize}
    \item In TDEs, a significant fraction of the total energy is radiated at late times. This is because the luminosity is observed to follow the mass fallback rate of stellar debris onto the SMBH. Estimates of the energy radiated during initial observations constitute $\lesssim 50\%$ of the total estimated radiated energy for the majority of the events in our sample (see Table~\ref{table:Lptp}).
    \item The late time luminosity ($\gtrsim 10$ peak timescales) extrapolated by assuming the luminosity continues to follow the fallback rate is consistent with the blackbody bolometric luminosities estimated from observations in \citet{van-Velzen:2019a} for 8 out of the 9 events in both samples, but the associated uncertainties are extremely large (see Figure~\ref{fig:Lobs_Lpred}). In their work, \citet{van-Velzen:2019a} argue that in order to explain the FUV luminosity, the total energy released at late times is significantly larger than the blackbody estimate suggests. If this is proven true with X-ray observations, our conclusion that the energy estimated during UV/optical observations is $\lesssim 50\%$ of the total radiated energy is  highly conservative. This would mean that the corresponding radiative  efficiencies  need to be larger than those estimated here.  
    \item The efficiency ($\epsilon$) of conversion from mass to radiated energy is somewhat degenerate with the mass of the disrupted star  (see Figure~\ref{fig:eff_mstar}). Calculating this efficiency requires untangling the amount of mass feeding the black hole from the luminosity of the event. The mass fallback rates used in the \mosfit model are self-similar over large ranges of stellar masses \citep{Mockler:2019a}, encumbering our attempts to isolate the effects of the stellar mass from those of the efficiency on the observed light curve. 
    \item The efficiency ($\epsilon$) of conversion from mass to radiated energy for tidal disruption flares are  consistent with AGN efficiencies and, in most cases, with stream collision efficiencies (see Figure~\ref{fig:DavisLaor}). 
    \item We find that near the peak of the transient light curve, the mass fallback rate is often close to, or greater than, the Eddington mass fallback rate, resulting in peak luminosities that are a significant fraction of the Eddington limits of the black holes (see Figure~\ref{fig:LpLedd_Mh}). This often requires that the efficiency is reduced near the peak of the light curve (see \citet{Dai:2018} for simulation work addressing this efficiency `suppression'). Therefore estimates of the efficiency comparing the peak bolometric luminosity to theoretical mass fallback rates will likely under-predict the average value of the efficiency. 

\end{itemize}

We have shown that for a TDE with a well-resolved light curve permit a significant reduction of the number of potential combinations of star and SMBH properties. However, when attempting to estimate  the conversion efficiency  from mass to electromagnetic radiation from a  well-resolved light curve, the largest  model uncertainty in our measurement comes from the degeneracy between the efficiency and the mass of the disrupted star. Our current mass fallback rates rely on polytropic solutions to stellar density profiles. An important step towards improving our stellar mass estimate is to use mass fallback curves that are based on more accurate stellar structures.
There have been several promising developments in this area, as multiple groups have recently produced simulations of disruptions using realistic (MESA) stellar profiles. Although none of these studies vary over the full impact parameter range of the disrupted stars (as was done in \citet{Guillochon:2013a} using polytropes), they already highlighting some  measurable differences in the shape of the fallback rate curve, particularly for higher mass ($3 M_{\odot}$) stars \citep{Law-Smith:2019, Golightly:2019b, Ryu:2019a}. It is our hope that incorporating this information into our models in future work will make it easier to constrain the mass of the disrupted star, and thus the associated radiative efficiency. 
Interestingly, \citet{Law-Smith:2019} have shown that the composition of the debris that falls back to the black hole changes with time and is dependent on the mass and age of the star. It is thus of paramount importance  to improve our understanding on how these composition anomalies might imprint themselves on the observed TDE spectra, so that we can better constrain the stellar mass of the disrupted star. This will enable a better characterization of both the mass to energy efficiency and the properties of the nuclear star clusters that surround the disrupting SMBHs.

\acknowledgments
We are indebted to J. Guillochon and K. Auchettl for their collaboration and for thier illuminating perspectives on many of the ideas included in this work. Our thoughts on the topics discussed here have been clarified through useful exchanges with J. Dai, J. Law-Smith, D. Kasen and N. Roth. E.R.-R. and B.M  are grateful for support from the Packard Foundation, The Danish National Research Foundation (DNRF132), NASA ATP grant NNX14AH37G and NSF grants (AST-161588, AST-1911206 and AST-1852393). The calculations for this research were carried out in part on the UCSC supercomputer Hyades, which is supported by National Science Foundation (award number AST-1229745) and UCSC. Calculations for this work also made use of an HPC facility funded by a grant from VILLUM FONDEN (project number 16599) and affiliated with the University of Copenhagen.

\bibliographystyle{yahapj}
\bibliography{library}

\end{document}